\documentstyle[aps,twocolumn,epsf]{revtex}

\begin{document}

\twocolumn[\hsize\textwidth\columnwidth\hsize\csname @twocolumnfalse\endcsname

\draft
\title{Magnetic excitation spectrum of dimerized antiferromagnetic
chains}

\author{G\"otz S.~Uhrig and H.J.~Schulz}

\address{Laboratoire de Physique des Solides, Universit\'e Paris-Sud, 
91405 Orsay, France}

\date{\today}

\maketitle

\begin{abstract}
Motivated by recent measurements on CuGeO$_3$ the spectrum of magnetic
excitations of an antiferromagnetic $S=\frac{1}{2}$ chain with alternating
coupling strength is investigated.  Wave vector dependent magnons and a
continuum with square root behavior at the band edges are found. The
spectral density of the continua is calculated.  Spin rotation symmetry
fixes the gap of the continuum to be twice the elementary magnon gap. This
is in excellent agreement with experimental results.  In addition, the
existence of bound states of two magnons is predicted: below the continuum a
singlet and a triplet, above the continuum an ``anti-bound'' quintuplet.
The results are based on field theoretic arguments, RPA calculations, and
consideration of the limit of strong alternation.
\end{abstract}
]

One--dimensional quantum antiferromagnets can exhibit the interesting
phenomenon of the spin--Peierls transition, i.e. a spontaneous dimerization
of the lattice and the exchange constant. The recent discovery
\cite{hase93a,nishi94,pouge94,hirot94} of a spin--Peierls transition in the
inorganic compound CuGeO$_3$ for the first time made neutron scattering
investigations in the ordered state possible.  These provide interesting
information on the dynamic spin correlations and therefore on the magnetic
excitations.  

The neutron scattering experiments exhibit a number of interesting features:
the magnon line shape asymmetry \cite{regna96a,marti96a}
 observed experimentally and
the interpretation of the behavior of the equal-time structure factor
\cite{regna96a} indicate quite clearly the existence of a continuum besides a
dominant magnon peak.  Its weight is roughly equivalent to the one in
the main peak \cite{regna96a}.  Very recent neutron scattering measurements
with high resolution \cite{ain96} show a ``double gap'' structure: at
$k=\pi$ one finds beside a magnon peak at the gap $\omega=\Delta$ another
gap which separates the magnon from the band edge of the continuum. It turns
out that this second gap equals within experimental accuracy the first one,
i.e.\ the onset of the continuum is at $2\Delta$.  Furthermore, there is
presently an active debate on the importance of next-nearest neighbor
couplings \cite{riera95,casti95,haas95}.  Evidence for a finite competing
next-nearest neighbor coupling stems from fits of the magnetic
susceptibility\cite{riera95,casti95} $\chi(T)$ and from a fit of the
equal-time structure factor.

These points motivate the present work. Its aim is to clarify theoretically
the magnetic excitation spectrum in the dimerized, spin-Peierls phase below
the transition temperature $T_{\scriptstyle\rm SP} \approx 14.3$K in the
dimerized phase.

In the dimerized phase it is reasonable to adopt an adiabatic approach
toward the coupling between the magnetic and the phonon subsystem
\cite{riera95,casti95,haas95,tsvel92}.  This is possible since the phonon
subsystem is three dimensional and hence its fluctuations are weak. This
holds in particular well below the transition temperature.  It seems
furthermore legitimate to neglect to first approximation the interchain
coupling\cite{nishi94,regna96a}.

For the above reasons we will focus on the following Hamiltonian
\begin{equation}
\hat H = J\sum_i [  (1+(-1)^i \delta) \bbox{S}_i\cdot \bbox{S}_{i+1}
+ \alpha \bbox{S}_i\cdot \bbox{S}_{i+2} ] ,
\label{hamilton1}
\end{equation}
where the $\bbox{S}_i$ are spin--1/2 operators.  Beside nearest-neighbor
coupling $J>0$, next--nearest neighbor coupling $\alpha J$ has been
included. The values of $\alpha$ used so far range between $0.24$ and $0.36$
\cite{riera95,casti95,haas95}. We now go to a fermionic representation of
the physics of the model via the Jordan--Wigner transformation
\cite{jor_tran} to fermionic degrees of freedom described by operators
$c_i$, $c^\dagger_i$. One then has $S^z_i= n_i -1/2$, $n_i = c^\dagger_ic_i$, and $S^+_i
= c_i^\dagger \prod_{j=-\infty}^i (1-2 n_i)$, and for $\alpha=0$ one obtains
\begin{eqnarray}
\label{eq:ham2}
\!\!\! \hat H&=& \frac{J}{2}\sum_i
(1+(-1)^i \delta) \\
\nonumber
& \times &
[- c^\dagger_i c^{\phantom{+}}_{i+1}
- c^\dagger_{i+1} c^{\phantom{+}}_{i}
+ 2(\frac{1}{2}- n_i)(\frac{1}{2}- n_{i+1})] .
\end{eqnarray}
For $\alpha \neq 0$ next--nearest neighbor terms are also present. For
$\delta = 0$ the properties of this model are well--understood, mainly
thanks to Bethe's exact
solution \cite{bethe_xxx,faddeev_spin1/2,luther_chaine_xxz,haldane_xxzchain}.
In particular, there are no spin--wave excitations, but the low--lying
excitations rather form a spinon continuum with lower bound $(\pi J/2) \sin
k$.

We now note that in the dimerized phase the momentum $k$ is not conserved:
any mode at $k$ is coupled to the mode $k+\pi$. If $\omega$ belongs to the
spectrum at $k$, it does so at $k+\pi$ as well. Together with the inversion
symmetry $k \leftrightarrow -k$ one obtains thus reflection symmetry about
$k=\pi/2$.  This symmetry is not taken into account in previous results
\cite{haas95,tsvel92}.  The reflection symmetry, however, only refers to the
energies, not to the spectral weights.

We first consider the case of weak dimerization, $\delta \ll 1$. Then a
continuum approximation for the fermionic model (\ref{eq:ham2}) is
appropriate, and one obtains the Hamiltonian of the so--called massive
Thirring model, for which there exists an exact Bethe ansatz
solution\cite{bergknoff_mtm_exact}. The model describes relativistic
fermions with ``light velocity'' $v=\pi J/2$ and a bare mass $m_0 \propto
\delta$. In addition there is a fermion--fermion interaction of strength
$g$. The known scaling relation \cite{cross79} for the magnetic gap (the
renormalized mass in fermionic language), $\Delta \propto \delta^{2/3}$ (up
to logarithmic corrections) uniquely fixes $g=2v$.  The exact solution
provides valuable information about the spectrum: first there are
single--particle and single--hole excitations which in the spin language
correspond to states with $\Delta M_z=\pm 1$, with excitation energy
$\sqrt{\Delta^2+v^2k^2}$. From the fermionic standpoint it is then clear
that there is a particle--hole continuum ($\Delta M_z =0$) with minimum
energy $2 \Delta$. However, due to the presence of interactions in the
model, there is a well--defined exciton--like particle--hole bound state
below this continuum\cite{bergknoff_mtm_exact}, at energy $\Delta$ when
$g=2v$. The particle, hole, and exciton thus form the low--energy degenerate
triplet excitation (the {\em magnon}) expected because of spin rotation
invariance of the original lattice model\cite{haldane}. An applied magnetic
field would split this triplet into its three components. In addition, there
is another excitonic bound state at energy
$\sqrt{3}\Delta/2$\cite{bergknoff_mtm_exact}. This, however, has no
counterpart in the $\Delta M_z=\pm 1$ sectors, and therefore is a singlet
excitation. Contrary to what was proposed in ref.\cite{tsvel92} it is thus
not possible to continuously connect (in $k$--space) the two bound state
excitations. We note that in a bosonized picture of the fermionic
theory\cite{luther_chaine_xxz,haldane_xxzchain}, the particle, hole, and
exciton correspond to soliton, antisoliton and breather modes. Also,
including a finite $\alpha$ affects the values of $v$ and $\Delta$
quantitatively, but does not affect the general structure of the spectrum
which is protected by spin rotation invariance.

From the elementary triplet at energy $\Delta$ it is possible to create
multiple excitation continua. Exciting two of the triplet modes, nine
different types of states can be created which form states with total spin
$S=0$, $1$, and $2$ and minimal energy $2\Delta$. Only the $S=1$ state is of
course accessible in a low--temperature neutron scattering experiment.  It
is interesting to consider what happens when the dimerization and therefore
the gap go to zero. Then $n$--magnon continua, with lower edge at $n\Delta$,
become close to the $2$--magnon continuum and they all coincide when
$\Delta\rightarrow 0$, giving rise to the well--known \cite{faddeev_spin1/2}
spinon continuum of the homogeneous spin--1/2 antiferromagnet. One should
however notice that the elementary constituents of the continua are $S=1$
objects when $\delta \neq 0$, whereas they are $S=1/2$ spinons (or
delocalized domain walls) for $\delta=0$. In fact, from the bosonic
representation \cite{luther_chaine_xxz,haldane_xxzchain} it is clear that a
nonzero $\delta$ acts as a linear confining potential between two spinons,
i.e. spinons can not exist as elementary excitations when $\delta\neq 0$.

From the above discussion the experimental observation of a $S=1$ continuum
at twice the energy of the elementary magnon \cite{ain96} seems very
natural. To investigate the spectral shape of the continuum further we
undertook a RPA study at zero temperature of the longitudinal spin
correlation function (the density correlation function in fermionic
language), starting from eq.(\ref{eq:ham2}). The general picture is the same
as that of the continuum limit: there is a gap $\Delta$ in the
single--particle(--hole) excitation, a particle--hole continuum above
$2\Delta$, and an excitonic bound state within this gap. The existence of
the gap for $\delta>0$ helps the approximation by providing an infrared
cutoff, i.e.\ by suppressing low--energy particle-hole excitations.  For
isolated dimers ($\delta\to 1, \alpha =0$), the RPA is exact. Yet in general
the approximation breaks the spin rotational invariance since the $S^z$
component is dealt with differently than the $S^x$ and $S^y$ components. In
a heuristic approach, this can be remedied by choosing an effective
interaction constant $J\to s J$ such that the exciton (magnon) is in the
middle of the two-particle continuum gap.  Since the static Fock diagram
renormalizes the hopping one works in addition with an effective energy
scale $J_{\scriptstyle\rm eff}$ and an effective alternation
$\delta_{\scriptstyle\rm eff}$. They are both enhanced in comparison to the
bare ones.
\begin{figure}[hbt]
\vspace*{0.5cm}
\centerline{\hspace*{2cm} \epsfxsize 7cm
{\epsffile{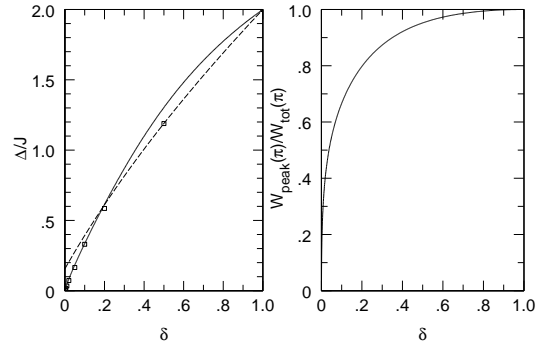}}
}
\vspace*{-0.5cm}
\caption[to]{
Left:
Gap $\Delta$ vs.\ alternation $\delta$ compared to perturbative
\protect\cite{harri73}
 (dashed line) and numerical results  (circles) 
 \protect\cite{soos85} at $\alpha=0$. Right:
relative magnon weight at $k=\pi$ vs.\ $\delta$ at $\alpha=0$.
}
\label{fi:ffig1}
\end{figure}
In fig.\ \ref{fi:ffig1}, results are shown for the alternation dependence of
the gap $\Delta(\delta)$ and the weight of the magnon $W_{\scriptstyle\rm
peak}(k)$ at $k=\pi$ relative to the total weight, i.e.\ peak plus
continuum.  The results at $\alpha=0$ are compared to the perturbation
results of Harris \cite{harri73} and numerical data of Soos {\it et al}.
\cite{soos85}. The agreement in the region of interest ($\delta \approx 0.03
- 0.06$ for CuGeO$_3$) is reassuring to proceed along this line.  Only for
very low $\delta<0.01$ our results become incompatible with the the
asymptotics\cite{cross79,spron86} $\Delta\propto
\delta^{2/3}/\sqrt{|\ln(\delta)|}$.  The right part of fig.\ \ref{fi:ffig1}
shows the spectral weight, which at $\delta=0$ is completely in the
diverging continuum\cite{luther_chaine_xxz,mulle81}.  It is shifted very
quickly into the magnon peak on dimerization.

\begin{figure}[hbt]
\vspace*{0.5cm}
\centerline{\hspace*{2cm}\epsfxsize 6cm
{\epsffile{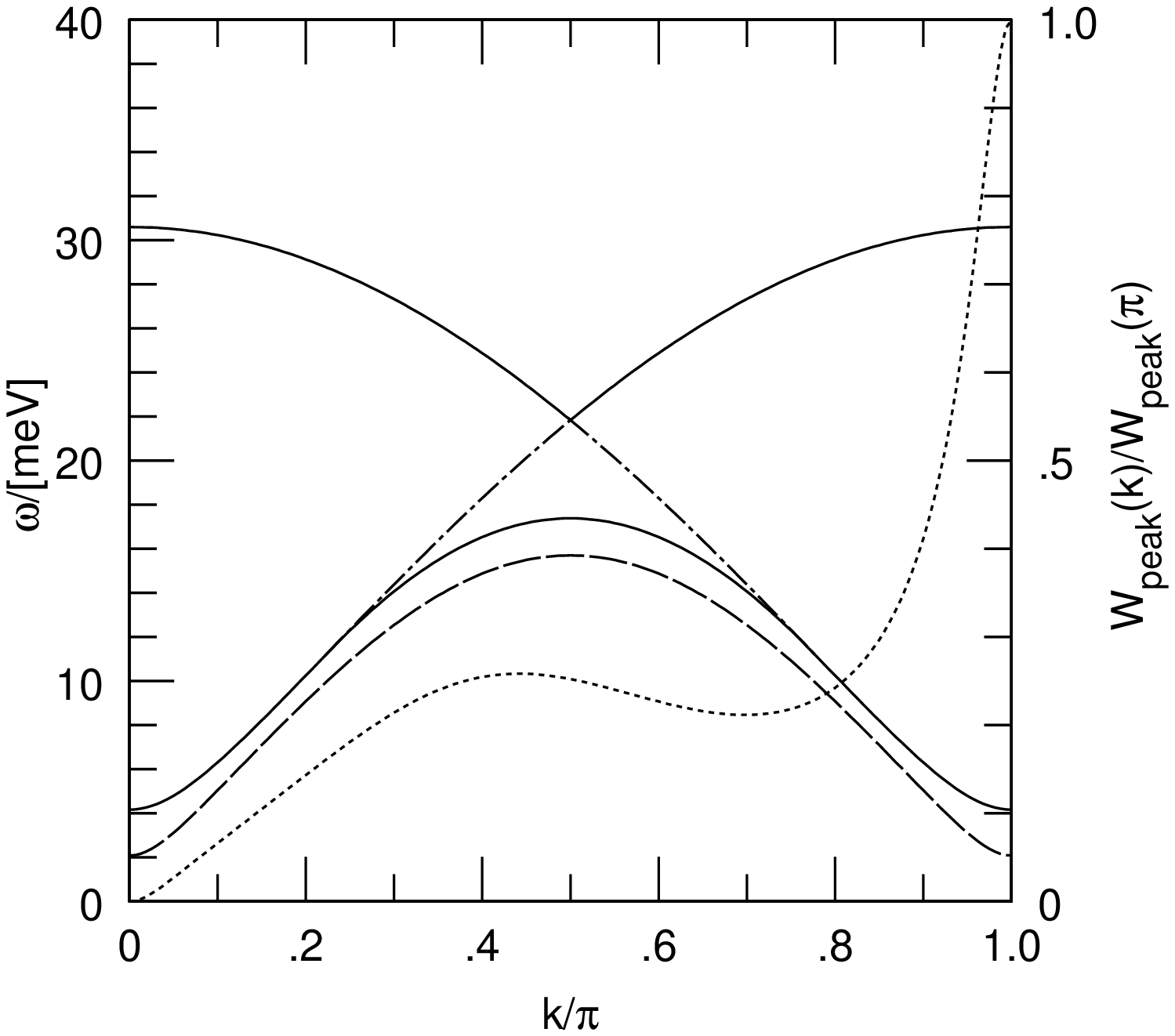}}
}
\vspace*{-0.5cm}
\caption{\label{fi:ffig2}
Dashed curve: magnon dispersion; dotted curve: relative magnon weight
 (right scale); continuum between solid lines; dashed-dotted lines:
 intermediate singularity. Numerical values:
$\delta_{\scriptstyle\rm eff}=0.136, J_{\scriptstyle\rm eff}=15.3\,$meV
 corresponding to $\delta=0.0506, J=11.2\,$meV.
}
\end{figure}
In fig.\ \ref{fi:ffig2}, the RPA spectrum is presented.  The long-dashed line
is the magnon dispersion.  The magnon weight relative to its value at
$k=\pi$ is depicted by the dotted line. Note its non-monotonic behavior.
The continuum is found between the two solid lines.  For $k<\pi/2$
noticeable continuum weight is found only between the lower solid line and
the dashed-dotted line.  The dashed-dotted lines show the positions of a
singularity between the band edges. It results from the local maximum at
$q=\pi/2$ of the particle-hole energy $\varepsilon_{\scriptstyle\rm ph}(q)
=\varepsilon(k/2+q)+\varepsilon(k/2-q)$ where $\varepsilon(q)=
J_{\scriptstyle\rm eff} \sqrt{\delta_{\scriptstyle\rm eff}^2 \sin^2(q) +
\cos^2(q)}$.  This maximum is not the global one for all $k$ satisfying
$\cos(k)<(1-\delta_{\scriptstyle\rm eff})/(1+\delta_{\scriptstyle\rm eff})$.

\begin{figure}[hbt]
\vspace*{0.5cm}
\centerline{\hspace*{2cm} \epsfxsize 7cm
{\epsffile{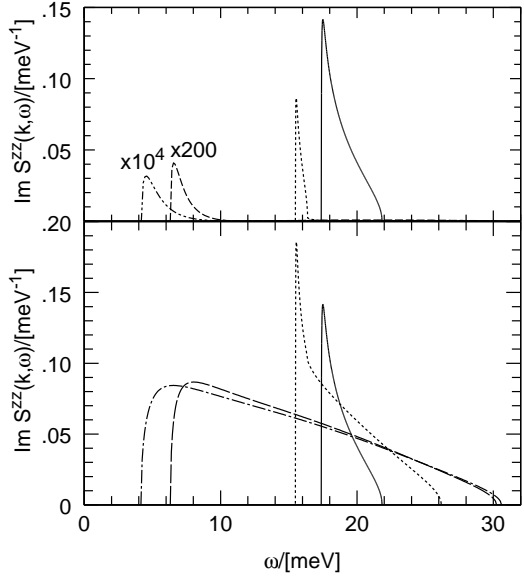}}
}
\vspace*{-0.5cm}
\caption{\label{fi:ffig3}
Continua for different wave vectors, the magnon $\delta$-peaks are not shown.
From left to right: upper part
$k/\pi=0.01, 0.1, 0.35, 0.5$; lower part $k/\pi=1.0, 0.9, 0.65, 0.5$.}
\end{figure}
The spectral densities are shown in fig.\ \ref{fi:ffig3}.  All singularities
(band edges and intermediate) are of square root type.  The continua at low
$k$ are extremely weak, but extend rigorously speaking up to the same high
energies as do their counterparts at $\pi-k$. For $k=(0.5\pm 0.15)\pi$ the
intermediate singularity (cf.\ fig.\ \ref{fi:ffig2}) is clearly seen at
$\omega\approx 17$meV. While for $k>\pi/2$ there is still considerable
weight above the intermediate singularity the spectral density for $k<\pi/2$
practically vanishes.  The extreme asymmetry of the continua at $k=\pi/2$
and $k=(0.5 \pm 0.15)\pi$ can be viewed as the precursor of another sharp
resonance which separates from the lower continuum edge in RPA for slightly
larger alternation values ($\delta > 0.082$). This is a bound state of two
magnons.

The physical origin of this second peak can be understood best in the limit
$\delta\to 1$.  Defining $\lambda=(1-\delta)/(1+\delta)$ and setting
$J(1+\delta)=1$ a perturbation expansion around complete dimerization can be
performed. To assess the influence of the NNN coupling $\bar\alpha=
\alpha/\lambda$ is kept constant so that $\lambda=0$ still corresponds to
isolated dimers.  These have $k$-independent triplet excitations at
$\omega=1$. The corresponding resolvent $R(k,\omega)=\langle S^z(k) (\omega
- L)^{-1} S^z(k) \rangle$ is $(1-\cos(k))/(4(\omega-1))$.  Here $L$ is the
Liouville operator\cite{fulde93}, which stands for commutation with $\hat
H$. For $\lambda>0$ this mode acquires a finite dispersion since the
triplets can hop from dimer to dimer with amplitude $-1/4$. In leading order
the magnon peak becomes $R_{\scriptstyle\rm peak}(k,\omega) =
p_0/(\omega-\omega(k))$ with
$4p_0=1-\cos(k)-\lambda(\cos(k)+2\bar\alpha\cos(2k))/4$ and
$\omega(k)=1-\lambda(1-2\bar\alpha)\cos(2k)/2$. For $\bar\alpha=0$, Harris
calculated $\omega(k)$ in third order\cite{harri73} in $\lambda$.  Note that
$\bar\alpha > 0$ reduces the triplet hopping.

A finite $\lambda$ couples the one-triplet subspace also to the $S=1$
two-triplet subspace. This leads in order $\lambda^2$ to the continuum.  By
a Mori-Zwanzig calculation\cite{fulde93} the corresponding spectral density
is found in leading order to be
\begin{equation}
\label{contex}
R_{\scriptstyle\rm cont}(k,\omega) = 
\frac{\lambda}{4}\frac{(1-\cos(k))(1-2\bar\alpha\cos(k))^2}
{y+a_1 - \mbox{sgn}(y)\sqrt{y^2-t^2}}
\end{equation}
where $a_1=(1+2\bar\alpha)/2$, $t=(1-2\bar\alpha)|\cos(k)|$ and $y$ is the
rescaled energy $\omega=2+\lambda y$.  The spectral density is finite for
$y^2 < t^2$ where the square root in (\ref{contex}) becomes
$i\sqrt{t^2-y^2}$.  Thus the continuum is $\propto
\lambda\sqrt{t^2-y^2}/(t^2+a^2 +2ay)$.  This asymmetric form is
qualitatively similar to the results in fig.\ \ref{fi:ffig3}. The edge
singularities are square roots.  Since no phase transition is expected for
$\lambda\in(0,1)$ this supports the RPA result of square root behavior for
all finite alternation values in contrast to a previous prediction of a
divergence \cite{tsvel92}.

The positive value of $a_1$ is the signature of attraction between two
triplets.  A $S^z=0$ triplet couples to the linear combination of two
antiparallel $S^z=1$ and $S^z=-1$ triplets.  On adjacent dimers they gain
$\lambda a/2$ in energy.  For $a_1 > t$ a bound state emerges which is
favored by $\bar\alpha>0$.  Its energy is found at the zero of the
denominator of (\ref{contex}).  Since $t$ vanishes at $k=\pi/2$ the bound
state is most prominent at this value. For $\bar\alpha=0$ the bound state
exists for $k\in [\pi/3,2\pi/3]{}$; for $\bar\alpha>1/6$ it is present at
all $k$.

Similarly one can investigate the two-triplet dynamics in the $S=0, 2$
subspaces. The secular equation found equals the denominator of
(\ref{contex}) with interactions $a_2=-a_1$ for the quintuplet and
$a_0=2a_2$ for the singlet.  This implies an anti-bound quintuplet above the
continuum and a more strongly bound singlet below the triplet state for all
$k$ even at $\bar\alpha=0$. The singlet and the quintuplet
 where also found at
$\bar\alpha=0$ by Harris\cite{harri73}. 
The attraction for the triplet was missed presumably due to
a sign error. It is natural to identify this
singlet with the $\sqrt{3}\Delta/2$ singlet found in the continuum
description above.  At $\bar\alpha=0$ and $k=0,\pi$
 the singlet lies at the lower
continuum band edge which is confirmed numerically\cite{bonne82} for a wide
range of $\delta$.
\begin{figure}[hbt]
\vspace*{0.5cm}
\centerline{\hspace*{2cm} \epsfxsize 7cm
{\epsffile{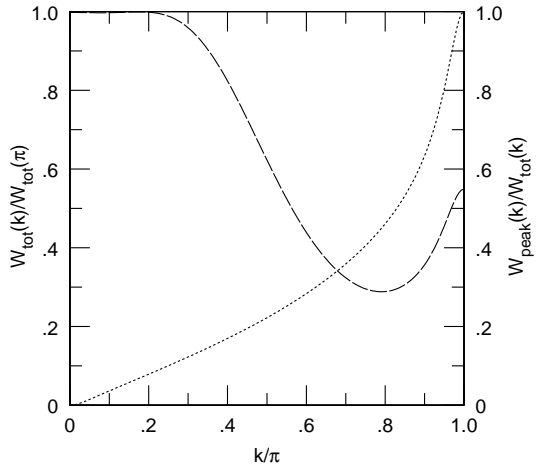}}
}
\vspace*{-0.5cm}
\caption{\label{fi:ffig4}
Dashed curve: magnon weight relative to total weight, right scale;
dotted curve: total weight normalized by its value at $k=\pi$, left scale.
}
\end{figure}
Reassured by the coherence of the results of the different approaches we
present in fig.\ \ref{fi:ffig4} the $k$ dependence found in RPA of
$W_{\scriptstyle\rm tot}(k)$, proportional to the equal-time structure
factor $S(k)$, and of $W_{\scriptstyle\rm peak}(k)$.  The agreement with the
experimental data\cite{regna96a} is very good. 
It was previously
stressed\cite{haas95} that $\alpha\approx1/4$ is necessary to achieve
agreement whereas the data in fig.\  \ref{fi:ffig4} is computed at
$\alpha=0$. This suggests
that $S(q)$ depends mainly on the gap $\Delta$ and
not so much on $\alpha$ if $\Delta$ is kept constant by tuning $\delta$.
But the interpretation requires care since the
approximation tends to raise $W_{\scriptstyle\rm tot}(k)$ around $\pi/2$.
Recall in this context the experimental fit result $\delta=0.042$
which is between our value ($\delta=0.0506$) and the one used so
far\cite{riera95,casti95,haas95}. 
The fact that for $k/\pi$ between $0.6$ and $0.9$ the spectral weight
is found mainly in the continuum is important for the
experimental determination of $S(k)$ \cite{regna96a}.

Summarizing, we give general arguments for the continuum onset at twice the
magnon gap in excellent agreement with experiment\cite{ain96}.  In RPA and
for $\delta\to 1$ we find a square root singularity at the continuum edge
which is also in agreement with experiment\cite{ain96}.  Detailed results for
the peak weights and the form of the continua are presented.  We find
evidence in the whole range of $\delta>0$ for a singlet excitation below the
continuum for all $k$.  In addition, we predict triplet (quintuplet)
excitations below (above) the continuum around $k=\pi/2$ for not too small
$\delta$. Their (anti)binding tendency is strongly enhanced by competing NNN
coupling.  An experimental identification of either of these excitations
would be particularly interesting.

We note that in ref.\cite{ain96} the continuum is attributed to solitonic
excitations of the dimerization pattern. However, one then has a important
deformation of the crystal lattice and would expect therefore a considerable
increase of the effective mass of these objects. The fact that the
experimental energy scale of the continuum is the bare $J$ favors a purely
magnetic origin for the continuum. Also, in the three--dimensionally ordered
structure below $T_{\rm SP}$ solitons on individual chains accompanied by
lattice deformations are confined by the interaction with adjacent chains
and therefore cannot be considered as free excitations.

The authors like to thank M.~A\"\i n, L.~P.~Regnault, 
S.~Haas, and E.~Dagotto for communicating their data.
This work was supported by the DFG
(GSU, individual grant and SFB 341) and by the EEC grant 
ERBCHRXCT 940438.



\end{document}